\documentclass[aps,prl,twocolumn,floatfix,showpacs,superscriptaddress]{revtex4}
\usepackage{epsfig}
\usepackage{bm}
\usepackage{latexsym}
 
\begin{document}

\title{A super-Ohmic energy absorption in driven quantum chaotic systems}
\author{A.~Ossipov}
\email{aossipov@ictp.trieste.it}
\affiliation{The Abdus Salam International Centre for Theoretical Physics,
Strada Costiera 11, 34014 Trieste, Italy}
\author{D.~M.~Basko}
\affiliation{The Abdus Salam International Centre for Theoretical Physics,
Strada Costiera 11, 34014 Trieste, Italy}
\author{V.~E.~Kravtsov}
\affiliation{The Abdus Salam International Centre for Theoretical Physics,
Strada Costiera 11, 34014 Trieste, Italy}
\affiliation{Landau Institute for Theoretical Physics,
2 Kosygina Street, 117940 Moscow, Russia}

\date{\today}

\begin{abstract}
We consider energy absorption by driven chaotic systems of the
symplectic symmetry class. According to our analytical
perturbative calculation, at the
initial stage of evolution the energy growth with time can be
faster than linear.
This appears to be an analog of weak anti-localization in
disordered systems with spin-orbit interaction.
Our analytical result is also confirmed by numerical calculations for
the symplectic quantum kicked rotor.
\end{abstract}

\maketitle

{\it Introduction.}---
The problem of energy absorption in a system driven by an external
time-dependent field is fundamental and important in many areas of
modern physics.
For a metallic sample of the volume~${\cal V}$ in an external
electric field $E(t)=E_{0}\cos\omega{t}$ the textbook solution
to this problem is given by the expression for the Joule heating
in the Ohmic regime: $W_0={\cal V}\,\sigma_{0}\,E_{0}^{2}/2$,
is the {\em constant} energy absorption rate determined by the
Drude conductivity~$\sigma_0$.
%\begin{equation}
%\label{Jh}
%W_{0}={\cal V}\,\sigma_{0}\,E_{0}^{2},
%\end{equation}
%where $W_{0}$ is the {\em constant} energy absorption rate
%in a sample of the volume~${\cal V}$, $E_{0}$ is the amplitude
%of the external  electric field $E(t)=E_{0}\cos\omega{t}$ and
%$\sigma_{0}$~is the  Drude conductivity.

This classical picture is based on the linear response theory for
systems with essentially continuous spectrum of electron states.
For quantum systems with few  degrees of freedom or for mesoscopic
systems with many degrees of freedom but still appreciable level
separation the Ohmic regime may break down leading to
a {\it time-dependent} absorption rate~$W(t)$. 

This point can be illustrated by an example of the quantum kicked
rotor (QKR) with the Hamiltonian:
\begin{equation} 
\label{QKR}
\hat{H}=\frac{\hat\ell^2}{2I}+K\,V(\theta)
\sum_{n=-\infty}^{\infty}\delta(t-nT),\quad
V(\theta)=\cos\theta,
\end{equation}
where $\hat\ell=-i\partial/\partial\theta$ is the angular
momentum, $I$~is the moment of inertia, and $K$~is a constant
controlling the strength of perturbation. For generic sufficiently
large~$K$ the classical dynamics described by the
Hamiltonian~(\ref{QKR}) is completely chaotic.
The period-averaged  energy absorption rate $W_0=K^2/(4TI)$
in this case is independent of time, analogously to the Ohmic
absorption.
Yet at sufficiently long times $t\gg t_{*}\sim K^{2}I^{2}/T$
the Ohmic regime breaks down because of the accumulation of
quantum corrections and the absorption rate decreases to zero.
This effect is known as {\it dynamic localization}
(DL) in the energy space~\cite{CCFI79}, and is analogous to
Anderson localization for disordered systems~\cite{Fishman}.
Such behavior is not specific to QKR, it occurs in other
chaotic systems~\cite{Stokmann,Haake}.

However, if $T/(4\pi I)$ takes a rational value,
the separation between certain energy levels of the rotor becomes
an integer multiple of the frequency~$2\pi/T$, and the absorption
rate is linear in time: $W(t)\propto t$~\cite{CCFI79}.
The same takes place for a harmonic oscillator coupled to the
external harmonic field via the coordinate, when the frequency
of the field is exactly at resonance with the oscillator
frequency~\cite{Husimi}. Fermi accelerator is another example
of a system where $W(t)$~can grow with time~\cite{FermiAcc}.
Such an anomalous ({\it growing} with time) {\it super-Ohmic}
behavior is typical of resonances. In contrast to the localization,
it is analogous to the {\em ballistic} transport through resonant
levels in a tight-binding model of 1d crystals. One can trace it
back to the classical integrability of the system with a
time-dependent perturbation.

In this letter we consider a class of {\it chaotic} systems
without resonances which show the super-Ohmic energy absorption.
Namely, we focus on the {\it quantum corrections} $\delta W(t)$
to the energy absorption rate in the time-dependent random matrix
theory (RMT) of the {\it symplectic} symmetry class, described by
the Hamiltonian:
\begin{equation}
\label{RMT}
\hat{H}(t)=\hat{H}_0+\hat{V}\phi(t),
\end{equation}
which possesses the time-reversal symmetry, but not the
spin-rotation symmetry.
Here $\hat{H}_0$~and~$\hat{V}$ are random matrices~\cite{M90}
whose symmetry will be specified below, and $\phi (t)$~is a given
function of time. This model describes
e.~g. the dynamics of electrons in driven quantum dots in the
presence of a spin-orbit interaction.

The corresponding problem for the orthogonal and unitary symmetry
classes has been recently considered~\cite{BSK03} and
{\it analytical} expressions for $\delta W(t)$ have been obtained.
For the harmonic perturbation $\phi(t)=\cos\omega t$ 
switched on at $t=0$, the absorption  rate~$W(t)$ appears to be
related to the frequency-dependent diffusion coefficient~$D(\omega)$
in a quasi-1d disordered wire of the corresponding symmetry class,
with the quantum corrections included:
\begin{equation}
\label{relat}
\frac{W(t)}{W_{0}}=\int_{-\infty}^{+\infty}\frac{d\omega}{2\pi}\,
\frac{e^{-i\omega{t}}}{(-i\omega+0)}\,\frac{D(\omega)}{D_{0}},
\end{equation} 
where $D_0$~is the classical diffusion coefficient.
This relationship does not contain any specific
feature of the model and is also valid for the QKR in the region
of parameters where it can be mapped onto the quasi-1d nonlinear
$\sigma$-model~\cite{AZ}. 
If Eq.~(\ref{relat}) is valid in the symplectic case as well, the
energy absorption rate $W(t)$ should {\it grow} with time beyond
the Ohmic limit $W_{0}$, as $D(\omega)$ is known to have
{\it positive} quantum corrections in the presence of a spin-orbit
interaction~\cite{LarHik}. Our calculations presented below show
that this is indeed the case.

{\it Choice of the model.}--- We adopt the following single-electron
Hamiltonians, which turns out to be the most convenient technically
for perturbative calculations:
\begin{equation}
\label{model2} 
\hat{H}(t)=\frac{\hat{\vec{p}}^2}{2m}+U(\vec{r})
+\hat{U}_{so}(\vec{r})+V(\vec{r})\,\phi(t), 
\end{equation} 
where $\hat{\vec{p}}=-i\vec\nabla$, and $U(\vec{r})$ and
$V(\vec{r})$ are independent Gaussian random fields: 
%\begin{equation} 
$\langle U(\vec{r})\,U(\vec{r}')\rangle=a_{U}\delta(\vec{r}-\vec{r}')$,
$\langle V(\vec{r})\,V(\vec{r}')\rangle=a_{V}\delta (\vec{r}-\vec{r}')$.
%\end{equation} 
The spin-orbit interaction is also taken to be
random~\cite{LarHik,Falko}:
\begin{equation}\label{Uso=}
\hat{U}_{so}(\vec{r})=
\vec{\sigma}\left[ \nabla U_{so}({\bf r})\times \hat{\vec{p}}\right],
\end{equation} 
where $\langle U_{so}(\vec{r})\,U_{so}(\vec{r}')\rangle=
a_{so}\delta(\vec{r}-\vec{r})$, and
$\vec{\sigma}=(\sigma^x,\sigma^y,\sigma^z)$ are Pauli matrices.

The advantage of the model~(\ref{model2}) is that the spin-orbit
coupling and the driving perturbation are represented by random
locally correlated fields. This makes it possible, after a proper
re-formulation~\cite{K03,YKK01} within the Keldysh
formalism~\cite{K64}, to apply basic rules of the impurity
diagrammatic technique~\cite{AGD} and its extension used in the
theory of weak Anderson localization and mesoscopic
phenomena~\cite{GLKh} to consider essentially nonlinear in the
driving perturbation, non-equilibrium problems. 

In the absence of the time-dependent term Eq.~(\ref{model2}) is
a basic model for describing disordered metals with a random
spin-orbit interaction. The kinetic energy term determines the
bulk density of states~$\nu$ (per unit volume, per spin
projection). Then $a_{U}=1/(2\pi\nu\tau_{0})$ and 
$\overline{[\vec{p}\times \vec{p}']^{2}}\,a_{so}=1/(2\pi\nu\tau_{so})$,
where $\tau_{0}$ is the momentum relaxation time,
$\tau_{so}\gg\tau_{0}$ is the spin relaxation time, and
$\overline{[\vec{p}\times\vec{p}']^{2}}$ denotes the momentum
product averaged over the Fermi surface. 
It has been shown~\cite{Efetov} that for a finite sample
in the long-time, low-energy limit $\epsilon\ll{1}/\tau_{so}\ll E_{Th}$
($E_{Th}$~being the Thouless energy) this model reduces to the
zero-dimensional nonlinear $\sigma$-model which, in turn, is equivalent
to the RMT of the symplectic symmetry class. This corresponds to
$\hat{H}_0$~in Eq.~(\ref{RMT}) being a random matrix from the Gaussian
symplectic ensemble (GSE) with the mean energy
level separation $\delta=1/(2\nu\mathcal{V})$.

The equivalence of the model~(\ref{model2}) to the time-depen\-dent
RMT can be also demonstrated, but it requires additional conditions:
\begin{equation}
\label{one-photon}
\frac{a_V}{a_U}=4\Gamma\tau_0\ll
\left(\frac{\omega}{E_{Th}}\right)^{2}\ll 1.
\end{equation}  
Here we have introduced a parameter $\Gamma=\pi\nu{a}_V/2$. We will
always be interested in the limit $\delta\ll\Gamma\ll\omega$, in which
case $1/\Gamma$~is the time required to absorb one photon,
as given by the Fermi Golden Rule.
The first condition~(\ref{one-photon}) allows to neglect the
multi-photon absorption processes which can easily violate the
condition of small energy transfer $\Delta\epsilon\ll E_{Th}$ even when
the condition $\omega\ll E_{Th}$ is fulfilled. Under the
conditions~(\ref{one-photon}) the last term of
Eq.~(\ref{model2}) corresponds to $\hat{V}\phi(t)$ of~Eq.~(\ref{RMT})
with $\hat{V}$~being a random matrix from the Gaussian {\it orthogonal}
ensemble (GOE), whose matrix element mean square is given by
$\Gamma\delta/\pi$.

We note that if instead of the time-dependent perturbation with
{\it random}~$V(\vec{r})$ one considers a more physical form of
the perturbation with some {\it fixed~$V(\vec{r})$} corresponding
to a uniform electric field or a modulation of the quantum dot
confinement potential, one can show~\cite{K03,YKK01} that under
the conditions~(\ref{one-photon}), where only the basic symmetries
(time-reversal, spin-rotation) matter, it will be equivalent to the
same RMT with a properly redefined~$\Gamma$.
However, calculations would be quite cumbersome in this case
because of the necessity to take into account boundary conditions.
The coupling by a random field $V(\vec{r})$ with local correlations
helps to avoid this technical problem.

We also note that instead of Eq.~(\ref{model2}) one can consider a
model where spin-orbit interaction is in the time-dependent term:
$\hat{H}(t)=\hat{\vec{p}}^{2}/2m+U(\vec{r})
+\hat{U}_{so}(\vec{r})\,\phi(t)$.
It corresponds to a different kind of the time-dependent
RMT, where in Eq.~(\ref{RMT}) $\hat{H}_0$~is taken from GOE and
$\hat{V}$~from GSE.
Still, the result turns out to be the same as for the
model~(\ref{model2}), being remarkably robust and independent of
whether the  spin-rotational invariance is broken in the
time-independent or in the time-dependent term [see also the
discussion after Eq.~(\ref{Cgoe})].

\begin{figure}[t]
\epsfig{figure=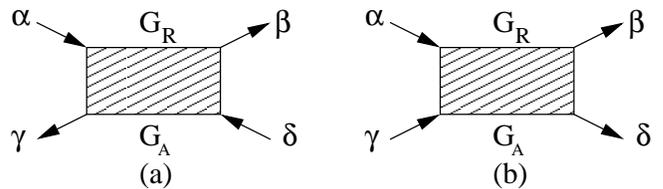,clip=,scale=1,angle=0}
\caption{\label{diff-coop} Diagrammatic representations for (a) 
diffuson, (b) cooperon; the Greek indices label the
spin projections $\uparrow,\downarrow$.}
\end{figure}

{\it Weak dynamic antilocalization.}--- The model~(\ref{model2})
allows for a  perturbative treatment which is very similar to the
theory of weak Anderson localization~\cite{GLKh}. 
The building blocks of this theory are ladder diagrams: the
diffuson and the cooperon (Fig.~\ref{diff-coop}).
The latter can also be represented as the maximally crossed
(``fan'') series of diagrams. In general they are functions of
time and momentum.
However, in the RMT, or ergodic, limit the main contribution to
observables is done by the {\it zero momentum mode}.
In this limit the diffuson
${\cal D}^{\alpha\beta}_{\gamma\delta}(t,t';\eta)$ and the cooperon
${\cal C}^{\alpha\beta}_{\gamma\delta}(\eta,\eta';t)$ are given by:
%\begin{widetext} 
%\begin{eqnarray}
%\label{diff}
%&&{\cal D}^{\alpha \beta}_{\gamma \delta}(t,t';\eta) =
%(1/2)\,D(t,t';\eta)
%\left[\delta_{\alpha\gamma}\delta_{\delta\beta}
%+e^{-(4/3)(t-t')/\tau_{so}}
%\vec{\sigma}_{\alpha\gamma}\vec{\sigma}_{\delta\beta}\right],\\
%\label{Coop}
%&&{\cal C}^{\alpha \beta}_{\gamma\delta}(\eta,\eta';t)=
%(1/2)\,C(\eta,\eta';t)
%\left[\sigma_{\alpha\gamma}^y\sigma_{\delta\beta}^y+
%e^{-(2/3)(\eta-\eta')/\tau_{so}}\left(\delta_{\alpha\gamma}
%\delta_{\delta\beta}+\sigma_{\alpha\gamma}^x\sigma_{\delta\beta}^x+
%\sigma_{\alpha\gamma}^z\sigma_{\delta\beta}^z\right)\right],\\
%&&{\cal C}^{\alpha \beta}_{\gamma\delta}(\eta,\eta';t)=
%C(\eta,\eta';t)
%\left[\langle\alpha\gamma|0\rangle\langle{0}|\beta\delta\rangle+
%e^{-(2/3)(\eta-\eta')/\tau_{so}}
%\sum_i\langle\alpha\gamma|i\rangle\langle{i}|\beta\delta\rangle\right], 
%\end{eqnarray} 
%\end{widetext} 
\begin{eqnarray}
\label{diff}
{\cal D}^{\alpha \beta}_{\gamma\delta}(t,t';\eta) &=&
(1/2)\,D(t,t';\eta)\times\\ \nonumber&&\times
\left[\delta_{\alpha\gamma}\delta_{\delta\beta}
+e^{-\frac{4(t-t')}{3\tau_{so}}}
\vec{\sigma}_{\alpha\gamma}\vec{\sigma}_{\delta\beta}\right],\\
\label{Coop}
{\cal C}^{\alpha \beta}_{\gamma\delta}(\eta,\eta';t)&=&
C(\eta,\eta';t)\times\\ \nonumber&&\times
\left[\langle\alpha\gamma|\hat{P}_0|\beta\delta\rangle+
e^{-\frac{2(\eta-\eta')}{3\tau_{so}}}
\langle\alpha\gamma|\hat{P}_1|\beta\delta\rangle\right], 
\end{eqnarray} 
where $\hat{P}_{0,1}$~are the projectors on the subspaces with the
total spin $S=0,1$, respectively:
\begin{eqnarray}
\langle\alpha\gamma|\hat{P}_0|\beta\delta\rangle&=&
\sigma_{\alpha\gamma}^y\sigma_{\delta\beta}^y/2,\\
\langle\alpha\gamma|\hat{P}_1|\beta\delta\rangle&=&
\left(\delta_{\alpha\gamma}\delta_{\delta\beta}
+\sigma_{\alpha\gamma}^x\sigma_{\delta\beta}^x+
\sigma_{\alpha\gamma}^z\sigma_{\delta\beta}^z\right)/2.
\end{eqnarray}
$D(t,t';\eta)$ and
$C(\eta,\eta';t)$ are defined as follows: 
\begin{eqnarray}
\label{Dgoe} 
D(t,t';\eta)&=&\theta(t-t')\exp\left\{
-\int_{t'}^t\gamma(t'',\eta) dt''\right\},\\
\label{Cgoe} 
C(\eta,\eta';t)&=&\theta(\eta-\eta')
\exp\left\{-\frac{1}{2}\int_{\eta'}^\eta\gamma(t,\eta'')d\eta''\right\},
\end{eqnarray}
where $\gamma(t,\eta)$ is determined by the external field:
\begin{equation}
\gamma(t,\eta)\equiv\Gamma[\phi(t+\eta/2)-\phi(t-\eta/2)]^2 .
\end{equation}

Note that Eqs.~(\ref{diff},\ref{Coop}) retain crossover triplet
terms.
At times larger than~$\tau_{so}$ they decay exponentially, and
the results reduce to those for the RMT model~(\ref{RMT}) with
$\hat{H}_0$~from GSE, which corresponds to $\tau_{so}\rightarrow{0}$,
so that in this model the triplet term would be absent from the
very beginning. The RMT model~(\ref{RMT}) with $\hat{H}_0$~from
GOE and~$\hat{V}$~from GSE would correspond to a finite~$\tau_{so}$.

The diagrammatic technique in the Keldysh representation allows
to calculate the time- and energy-dependent electron distribution
function~$f(\epsilon,t)$, from which one deduces the energy
absorption rate:
\begin{equation}
W(t)=\frac{\partial}{\partial{t}}\int\epsilon\,f(\epsilon)\,d\epsilon.
\end{equation}
The expansion in the number of the diffuson or cooperon loops
corresponds to the expansion in the powers of the mean level
spacing~$\delta$, which is assumed to be the smallest energy
scale of our problem~\cite{BSK03,K03}.
The leading contribution is given by diagrams containing no diffuson
or cooperon loops; it corresponds to the Ohmic absorption with the
rate
\begin{equation}
W_0=\frac{2\Gamma}{\delta}\,\overline{(\partial_t \phi)^2},
\end{equation}
where the overline denotes the average over the period.
The next order correction to the Ohmic absorption rate is
obtained by taking into account one-loop diagrams:
\begin{eqnarray}\nonumber
W(t)&=&W_0+\frac{\Gamma}{\pi}\int_0^t
\left(3e^{\frac{-4\eta}{3\tau_{so}}}-1\right)
\times\\ && \times\,
\partial_t\phi (t)\,\partial_t\phi (t-\eta)\,
C(\eta,-\eta;t-\eta/2)\,d\eta.\label{W1=}
\end{eqnarray}
For $\phi(t)=\cos\omega{t}$ the long-time behavior
($t\gg{1}/\Gamma$) of the above expression takes
the following form:
\begin{equation}
\label{correc}
\frac{W(t)}{W_0}=1+
\left\{\begin{array}{ll}
-\sqrt{t/t_*},&t\ll\tau_{so},t_*,\\
(1/2)\sqrt{t/t_*},&\tau_{so}\ll{t}\ll{t}_*,
\end{array}\right.
\quad
t_\ast=\frac{\pi^3\Gamma}{2\delta^2}.
\end{equation}
These two limiting cases differ by the presence or absence of the
triplet contribution and correspond to the orthogonal or symplectic
symmetry classes.
Thus the weak localization correction to the classical absorption
rate in the symplectic case is {\it positive} and its magnitude is
half that for the orthogonal case.
Exactly the same holds for the weak localization correction
to the conductivity of a quasi-1d disordered wire~\cite{LarHik}, so
the relation~(\ref{relat}) is extended also to the symplectic case.
%symmetry class.

The perturbative result~(\ref{correc}) is valid only for times
$t\ll t_\ast$. At $t\gg{t}_*$ the relation~(\ref{relat}) suggests
to use the known results for a quasi-1d wire~\cite{Efetov}, where,
in spite of the positive sign of the first weak localization
correction, all states are localized, with the localization
length four times larger than in the orthogonal case.
Localization implies $D(\omega)\rightarrow{0}$ at
$\omega\rightarrow{0}$, for which Eq.~(\ref{relat}) gives
$W(t)\rightarrow{0}$ at $t\rightarrow\infty$.

{\it Quantum kicked rotor with spin.}---
The exact time dependence of the correction~(\ref{correc}) is
derived rigorously for the disordered model~(\ref{model2}). However,
one can expect that the increase of the energy absorption rate in
time is a general result valid for an ergodic dynamical system
possessing symplectic symmetry. In order to support this statement
we introduce a spin degree of freedom into the standard kicked
rotor model~\cite{S89}. Correspondingly, $V(\theta)$ in the
Hamiltonian~(\ref{QKR}) acquires a $2\times{2}$ matrix structure:
%\begin{widetext}
%\begin{equation}
%\hat{H}=  \frac{\hat\ell^2}{2I}
%+K\left[\cos(\alpha)\cos(\beta)\cos(\theta)
%+\frac{1}{2}\cos(\alpha)\sin(\beta)\sin(2\theta)\sigma^x
%+\sin(\alpha)\sin(\theta)\sigma^z\right]
%\end{equation}
%\end{widetext}
\begin{eqnarray}\nonumber
V(\theta)&=&\cos\alpha\cos\beta\cos\theta+\\
&&+\frac{\sigma^x}{2}\cos\alpha\sin\beta\sin 2\theta
+\sigma^z\sin\alpha\sin\theta,
\end{eqnarray}
where the parameters $\alpha$~and~$\beta$ allow to switch between
different symmetry classes. Note that the presence of both
$\sigma^x$~and~$\sigma^z$ terms with {\em different} dependencies
on~$\theta$ is essential, as otherwise the spin sector of the
Hamiltonian could be diagonalized by a global spin rotation.

\begin{figure}[t]
\epsfig{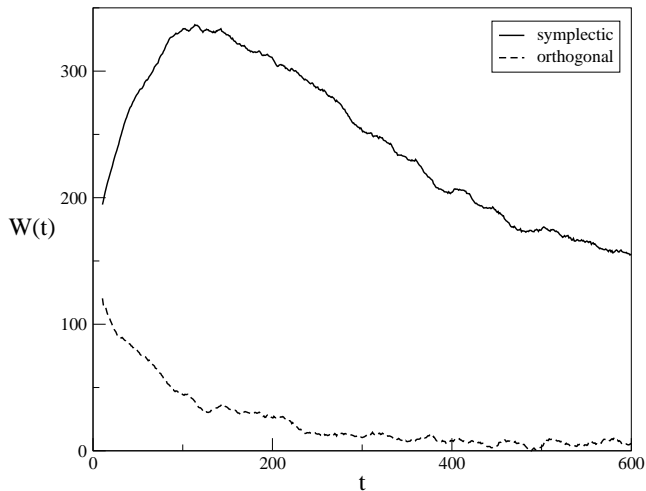}
\caption{\label{kr_rate} The energy absorption rate as a function of time for
  the standard and symplectic kicked rotor. The energy $E(t)$ is averaged over
  100 trajectories with different initial conditions.}
\end{figure}

In order to observe the dynamical anti-localization we fix the parameters
$\alpha =0.187$, $\beta= 1.284$ which corresponds to the symplectic symmetry
class~\cite{S89} and study the evolution of the wave-packets in the momentum
space. The other parameters are $T=1$, $I=10\,(\sqrt{5}-1)/(2\pi)$, $K=10\,I$,
and the Hilbert space size $\ell_{max}=16384$.
This model is very convenient for numerical
study since the application of the Floquet operator to a state  can be
performed by using the Fast Fourier Transform algorithm as for the standard
QKR model. Fig.~(\ref{kr_rate}) shows that the energy absorption rate
indeed initially increases in time until the strong localization changes this
behavior to the opposite one. For comparison we plot the energy absorption
rate for the standard QKR calculated for the same values of $T, I$ and $K$ as
well. Here, in contrast, the quantum correction is negative from the very
beginning.

{\em Conclusions.}--- The main results of the paper are represented by
Eq.~(\ref{correc}) and Fig.~\ref{kr_rate}. The former shows that the
analogy between the energy absorption by an ac driven chaotic quantum
dot and the propagation in a quasi-1d disordered wire is valid also in the
presence of the spin-orbit interaction, i.~e. for systems of the symplectic
symmetry class, at least at the level of the first weak anti-localization
correction. Numerical results for the quantum kicked rotor exhibit
qualitatively the same behavior: weak dynamic anti-localization at shorter
times, and strong dynamic localization at longer times.
At the weak anti-localization stage the energy for both chaotic systems
exhibits a peculiar super-Ohmic growth in time, in spite of the fact
that no resonances are present in the considered systems.

We thank V.~I.~Fal'ko for fruitful discussions. 
    
%******************************************************************************

\end{document}